\documentclass[12pt]{article}
\usepackage{amssymb}
\usepackage{amsmath}
\newtheorem{Definition}{Definition}

\newtheorem{Theorem}{Theorem}

\begin{document}

\begin{titlepage}

\title{Anatomy of Malicious Singularities}
\author{
Michael Heller, \\
Vatican Observatory, V-00120 Vatican City State;\footnote{The
correspondence address:  ul.  Powsta\'nc\'ow Warszawy 13/94, 33-110
Tarn\'ow, Poland. E-mail: mheller@wsd.tarnow.pl}\\
\and Zdzis{\l}aw Odrzyg\'o\'zd\'z, \\
\and Leszek Pysiak, \and
and Wies{\l}aw Sasin, \\
Technical University of
Warsaw,\\ Plac Politechniki 1, 00-661 Warsaw, Poland.}

\maketitle

\begin{abstract}
As well known, the b-boundaries of the closed Friedman world model
and of Schwarzschild solution consist of a single point. We study
this phenomenon in a broader context of differential and structured
spaces. We show that it is an equivalence relation $\rho $, defined
on the Cauchy completed total space $\bar{E}$ of the frame bundle
over a given space-time, that is responsible for this pathology. A
singularity is called malicious if the equivalence class $[p_0]$
related to the singularity remains in close contact with all other
equivalence classes, i.e., if $p_0 \in \mathrm{cl}[p]$ for every $p
\in E$. We formulate conditions for which such a situation occurs.
The differential structure of any space-time with malicious
singularities consists only of constant functions which means that,
from the topological point of view, everything collapses to a single
point. It was noncommutative geometry that was especially devised to
deal with such situations. A noncommutative algebra on $\bar{E}$,
which turns out to be a von Neumann algebra of random operators,
allows us to study probabilistic properties (in
a generalized sense) of malicious singularities. Our main result is
that, in the noncommutative regime, even the strongest singularities
are probabilistically irrelevant.
\end{abstract}

\date{}
\end{titlepage}\

\section{INTRODUCTION}

The classical period in the study of space-time singularities was closed by
the book by Hawking and Ellis \cite{HawkingEllis} and the extensive review
paper by Tipler, Clarke and Ellis \cite{TiplerRev}. It was Clarke \cite%
{Clarke} who later on summarized mathematical advancements in this field.
This does not mean that the interest in singularities has evaporated, still
many works appear in this area, but they are studying particular cases, or
classes of cases, rather than \textit{the problem} of singularities as such.
Moreover, a general feeling that the future theory of quantum gravity will
eliminate singularities from the physical image of the world also
contributes to the fact that to the majority of cosmologists and
astrophysicists the singularity problem seems to be less important than it
was several years ago. This prejudice is not necessarily true. First of all,
we still have no final quantum gravity theory, and we can only speculate
about the fate of singularity when such a theory will finally be discovered.
Moreover, it may happen, as we shall argue in this paper, that the study of
classical singularities can tell us something about the fundamental level of
physics and, in this way, contribute to our search for the quantum gravity
theory. And, last but not least, the singularity problem is highly
interesting from the mathematical point of view. Many branches of
mathematics are involved in this problem, such as: topology, measure theory,
noncommutative geometry, not to mention traditional differential geometric
methods in various its, often very subtle, modifications and improvements.
And it is highly probable that some new technics to deal with the problem
still await to be discovered.

In the present paper, we are interested in the structure of singularity and
its role in a physical theory rather than in computational details. There
are various kinds of singularities in general relativity and in similar
space-time theories \cite{Clarke75,EllisSchm}). As the subject-matter of the
present study we choose the strongest singularity of all that we call
malicious singularity (its definition is given in Sec. \ref{sec2}). The
initial and final singularities in the Friedman world models and central
singularity in the Schwarzschild solution belong to this kind. It was once
believed that their best mathematical description was provided by Schmidt's
b-boundary construction, but after ``pathologies'' had been discovered in
this construction by Bosshard and Johnson (see below) it slowly went into
oblivion. In the present paper, we try to convince the reader that the
crisis in the study of singularities was not due to the wrong approach
adopted by Schmidt, but rather by the malicious character of strong
singularities themselves. Moreover, by starting with Schmidt's construction
we hope to be able to throw a new light onto the singularity problem.

Let us first briefly recall Schmidt's construction \cite{Schmidt71}. Let $%
(M,g)$ be a space-time, and $\pi_M : E \rightarrow M$ a frame bundle over $M$
with the Lorentz group $G = SO(3,1)$ as its structural group. Levi-Civita
connection on $M$ determines the family of Riemann metrics on $E$. We select
one of them (noticing that the further construction does not depend on any
particular choice), with the help of it determine the distance function on $E
$ and, in the usual way, construct the Cauchy completion $\bar{E}$ of $E$.
Since the right action of the group $G$ on $E$ prolongs to that of $\bar{E}$%
, we define the quotient space $\bar{M} = \bar{E}/G$, and call $\bar{M}$ the
\textit{b-completion} of space-time $M$ ($M$ is open and dense in $\bar{M}$%
). Finally, we define the \textit{b-boundary} of space-time $M$ as $%
\partial_bM = \bar{M} - M$. Schmidt has demonstrated that every $g$%
-incomplete geodesic and every incomplete timelike curve of bounded
acceleration in $(M,g)$ (that can represent the history of a physical body)
determines a point in $\partial_bM$.

In his original work \cite{Schmidt71} Schmidt computed b-boundaries of a few
simplified space-times, and argued that his definition covers what was
intuitively regarded as singularities. However, when Bosshard \cite%
{Bosshard76} and Johnson \cite{Johnson77} computed b-boundaries for the
closed Friedman world model and for the Schwarzschild solution, it turned
out that the b-boundaries of these two space-times consisted of a single
point that was not Hausdorff separated from the rest of space-time. There
were several attempts to cure the situation, but new proposals were either
artificial or not general enough (for an extensive review see \cite{Dodson}).

In the first part of the present paper (Secs. \ref{sec1} and \ref{sec2}), we
identify singularities for which Schmidt's construction behaves
pathologically, and then investigate the cause of this behavior. We do this
with the help of slightly generalized but otherwise standard methods of
differential geometry [instead of differential manifold we use differential
spaces (Sec. \ref{sec1}) and structured spaces (Sec. \ref{sec2})] and
topology. It is shown that it is an equivalence relation $\rho \subset E \times E$
(essentially coming from the action of the group $G$ on $\bar{E}$) that is
responsible for the pathological behavior. If the equivalence class $[p_0]$
related to the singularity remains in a close contact with all equivalence
classes, i.e., if $p_0 \in \mathrm{cl}[p]$ for every $p\in E$, topological
pathologies occur, and the singularity is called \textit{malicious}. A
theorem is proved providing the equivalent conditions for this to happen.
This part of the paper extends results obtained in our former works \cite%
{Malicious,BanachCenter,GRG94,Structured,Acta92}.

It was noncommutative geometry that was especially devised to deal
with cases which go beyond possibilities of the traditional methods.
This is why, in the second part of the paper (Sect. \ref{sec3}), we
turn to noncommutative geometry for help. Every equivalence relation
gives rise to a groupoid, with the help of which one can define a
suitable (noncommutative) algebra that can be regarded as a
generalization of the algebra of smooth functions on a manifold
\cite[p. 86]{Connes}. In Subsec. \ref{ssec3.1}, we adapt this
strategy to the case of space-time with malicious singularities, and
in Subsec. \ref{ssec3.2}, we study, in the noncommutative setting,
probabilistic properties of malicious singularities.
In this case, the algebra modeling the noncommutative regime is a
von Neumann algebra of random operators. Owing to the corresponding
probability measure (in a generalized sense) even the strongest
singularities turn out to be irrelevant. They manifest themselves
only when one goes to the commutative regime.

If we agree that methods of noncommutative geometry will find their
application in the future quantum gravity theory then the results obtained
in the present paper acquire natural physical interpretation which we
briefly discuss in Sec. \ref{sec4}. In fact, the present work has been
inspired by a model based on noncommutative geometry proposed by us \cite%
{Finite,Observables,Random,JMP05,Pysiak06} and aimed at unifying general
relativity and quantum mechanics. However, the results of the present paper
are self-contained and do not depend of any particulars of this model.

\section{QUOTIENT DIFFERENTIAL SPACES}
\label{sec1}
Let us consider a family $C$ of real valued functions on a set $%
M$ endowed with the weakest topology $\tau_C$ in which the functions of $C$
are continuous. A function $f$, defined on $A\subset M,$ is called a \emph{%
local $C$-function\/} if, for any $x\in A, $ there exists a neighborhood $B$
of $x$ in the topological space $(A, \tau_A)$, where $\tau_ A$ is the
topology induced in $A$ by $\tau_C$, and a function $g\in C$ such that $%
g|B=f|B$. Let us denote by $C_A$ the set of all local $C$-functions. It can
be seen that $C\subset C_M$.

\begin{Definition}
If $C=C_M$, the family $C$ is said to be \emph{closed with respect to
localization}.
\end{Definition}

\begin{Definition}
$C$ is said to be \emph{closed with respect to superposition with }
$C^{\infty }$-\emph{functions\/ on a Euclidean space} if for any $n\in
\mathbf{N}$ and each function $\omega \in C^{\infty }(\mathbf{R}%
^{n}),\;f_{1},...,f_{n}\in C$ implies $\omega \circ (f_{1},...,f_{n})\in C$.
\end{Definition}

$C$ is obviously a commutative algebra \cite{Sikorski67}.

\begin{Definition}
A family $C$ of real valued functions on $M$ which is closed with respect to
localization and closed with respect to superposition with $C^{\infty }$-real valued
functions of several variables is called a \emph{differential structure\/} on $M$. A pair $(M,C)$%
, where $C$ is a differential structure on $M$, is called \emph{%
differential space}. Functions belonging to $C$ are called \emph{smooth
functions}.
\end{Definition}

Every differential manifold is a differential space with $C=C^{\infty}(M)$
as its differential structure.

Let $(M,C)$ be a differential space, and $\rho \subset M\times M$ an
equivalence relation in $M$. The family
\[
\bar{C}:=C/\rho =\{\bar{f}:\,M/\rho \rightarrow \mathbf{R}:\bar{f}\circ \pi
_{\rho }\in C\},
\]%
with $\pi _{\rho }:M\rightarrow M/\rho :=\bar{M}$ being the canonical
projection, is the largest differential structure on $M/\rho $ such that $%
\pi _{\rho }$ is smooth \cite{Sasin98,Walisz72}.

Let us define
\[
C_{\rho }:=\{f\in C:\forall _{x,y\in M}x\rho y\Rightarrow f(x)=f(y)\}.
\]%
Any function belonging to $C_{\rho }$ is called $\rho $-\emph{consistent}. $%
C_{\rho }$ is obviously a differential structure on $M$. The mapping $%
\pi _{\varrho }^{\ast }:C/\varrho \rightarrow C_{\varrho }$ is an
isomorphism of $\mathbf{R}$-algebras.

A subset $A \subset M$ is called {\emph{$\rho $-saturated} if together with
a point $x \in A$ it also contains its entire $\rho $-equivalence class $[x]$%
, i.e., if $\pi_{\rho }^{-1}(\pi_{\rho }(A))= A$. Let us denote the topology of
all open and $\rho $-saturated subsets by $\tau_{\rho }$. In
general, $\tau_{C_{\rho}}\subset \tau_{\rho }$. Let us observe that
\[
\tau_{C/\rho } = \tau_C/\rho \Leftrightarrow \tau_{C_{\rho }} = \tau_{\rho }.
\]
If one of these equivalent conditions is satisfied, the relation $\rho $ is
said to be \emph{regular}. If this does not hold, it
is reasonable to switch from differential algebras to sheaves of
differential algebras. (For more about quotient differential spaces see \cite%
{GRG94,Acta92}.) }

\section{QUOTIENT STRUCTURED SPACES}

\label{sec2}

\begin{Definition}
Let $(M,\tau )$ be a topological space. The sheaf $\mathcal{C}$ of real
functions on $(M,\tau )$ is said to be a \emph{differential structure} on $M$
if for any open set $U\in \tau $ and any set of functions $f_{1},\ldots
,f_{n}\in \mathcal{C}(U),\,\omega \in C^{\infty }(\mathbf{R}^{n})$, the
superposition $\omega \circ (f_{1},\ldots ,f_{1})\in \mathcal{C}(U)$. The
triple $(M,\tau ,\mathcal{C})$ is called the \emph{structured space}.
\end{Definition}

If $\tau $ is clear from the context, we shall also write $(M, \mathcal{C})$.

Any differential space can be trivially regarded as a structured space (let
us notice that the condition of the closeness with respect to localization
is already contained in the sheaf axioms). It can be shown that a structured
space $(M,\tau ,\mathcal{C})$ is a differential space if for every open set $%
U\in\tau$ and any point $x\in U$, there exists a function $\varphi$, called
\emph{bump function}, such that $\varphi (p)=1$ and $\varphi |M - U=0.$

In the following, we shall make use of structured spaces of \textit{constant
(differential) dimension}, i.e., the ones satisfying the following
conditions: (1) $\mathrm{dim}T_{p}M=n$ for every $p\in M$, (2) the module of
vector fields $\mathcal{X}(M)$ is locally free of rank $n$. (The theory of
structured spaces was developed in \cite{Structured}).

Let $\rho \subset M\times M$ be an equivalence relation in a structured space
$(M, \mathcal{C})$, and let us consider
the quotient topological space $(M/\rho ,\tau /\rho )$. The quotient
sheaf is given by
\[
\tau /\rho \ni U\mapsto (\mathcal{C}/\rho )(U)
\]%
where $(\mathcal{C}/\varrho )(U)=\mathcal{C}(\pi _{\varrho }^{-1}(U))/\varrho ,$ and we have the
quotient structured space $(M/\rho ,\mathcal{C} /\rho )$.

\textbf{Example 1} Let us consider a Euclidean plane $(\mathbf{R}%
^{2},C^{\infty }(\mathbf{R}^{2}))$ and the equivalence relation $\rho
\subset \mathbf{R}^{2}\times \mathbf{R}^{2}$ defined by $x\rho y$ iff $k\in
\mathbf{R}-\{0\}$ such that $x=ky$. The functions of $C^{\infty }(\mathbf{R}%
^{2})$, consistent with $\varrho $, are constant on equivalence classes,
i.e., $C_{\rho }^{\infty }(\mathbf{R}^{2})\simeq \mathbf{R}$, and everything
collapses to a point (in the sense that constant functions do not distinguish
points). Let us denote $\mathbf{R}^{2}/\rho $ by $\bar{M}$, the quotient sheaf by
$\bar{\mathcal{C}}$, and define the quotient
structured space $(\bar{M},\bar{\mathcal{C}})$.
The singular point $\{(0,0)\}$, which we shall denote by $x_{0}$, regarded
as an equivalence class, remains in the \textquotedblleft close
contact\textquotedblright\ with other equivalence classes, i.e., $x_{0}\in
\mathrm{cl}[x]$. The singular point $x_{0}$ is not Hausdorff separated from $%
\bar{M}$, and $\bar{M}$ is the only open set containing it. Let us introduce
the abbreviations: $\mathbf{R}^{2}-\{x_{0}\}=M$ and $\mathcal{C}(\mathbf{R}%
^{2}-\{x_{0}\})=\mathcal{C}(M)$. It is evident that the differential space $(M,%
\mathcal{C}(M))$ is a one-dimensional projective space $(\mathbf{P}^{1}(%
\mathbf{R}),C^{\infty }(\mathbf{P}^{1}(\mathbf{R}))$ which can be regarded
as a regular (i.e., non-singular) part of the structured space $(\bar{M},%
\bar{\mathcal{C}})$. Moreover $M$ is open and dense in $\bar{M}$,
and consequently $\{x_{0}\}=\bar{M}-M$ can be regarded as a singular
boundary of $M$.

The above example motivates the following definition.

\begin{Definition}
Let $\rho = E \times E$ be an equivalence relation in $E$. An equivalence
class $[p_0]$, $p \in E$, of this relation is a \emph{malicious singularity}
in $E/\rho $ if $p_0$ remains in a close contact with all equivalence
classes of the relation $\rho $, i.e., if $p_0 \in \mathrm{cl}[p]$ for every
$p \in E$.
\end{Definition}

Let us consider two structured spaces: $(E, \mathcal{C}_E)$ of constant
differential dimension $n$, and $(F, \mathcal{C}_F)$ of constant
differential dimension $k\leq n$.

\begin{Definition}
An equivalence relation $\rho \subset E \times E$ is said to be $F$\emph{%
-regular with the exception of points} $p_1, \ldots ,p_r$ if the points $%
p_i,\, i=1,\ldots ,r$, \emph{are fixed points of the relation} $\rho $\emph{%
, i.e.,} if $[p_i]=\{p_i\}$, and the relation $\rho \cap (E-K) \times (E-K)$%
, where $K = \{p_1, \ldots , p_r\}$, is $F$ regular in $E-K$, i.e., if for
every $p \in E-K$ there exists an open neighborhood $\pi_{\rho}^{-1}(U) \ni p
$, where $U \in \mathrm{top}(E-K)/\rho$, and a diffeomorphism $\Phi :
\pi_{\rho }^{-1}(U) \rightarrow U \times F$ such that the following diagram
commutes

\end{Definition}

\vspace{1.5cm}

\unitlength=0.70mm \special{em:linewidth 0.4pt} \linethickness{0.4pt}
\begin{picture}(100.00,120.00)
\put(43.00,128.00){\makebox(0,0)[cc]{$\pi_{\rho }^{-1}(U)$}}
\put(95.00,128.00){\makebox(0,0)[cc]{$U\times F$}}
\put(70.00,95.00){\makebox(0,0)[cc]{$U$}}
\put(55.00,128.00){\vector(1,0){30.00}}
\put(55.00,125.00){\vector(1,-2){13.00}}
\put(85.00,125.00){\vector(-1,-2){13.00}}
\put(70.00,132.00){\makebox(0,0)[cc]{$\Phi $}}
\put(55.00,110.00){\makebox(0,0)[cc]{$\pi_{\rho }$}}
\put(85.00,110.00){\makebox(0,0)[cc]{$\pi_{\rho }$}}
\end{picture}

\begin{Theorem}
\vspace{-6cm}
Let $(E, \mathcal{C}_E)$ be a structured space of constant dimension $n$, $%
(F, \mathcal{C}_F)$ a structured space of constant dimension $k\leq n$, and $%
\rho \subset E \times E $ an equivalence relation on $E$ that is $F$-regular
with the exception of points $p_1, \ldots , p_r \in E$. The following
conditions are equivalent:

\begin{enumerate}
\item the points $p_i, \,i=1,\ldots , r$, remain in close contact with all
equivalence classes of the relation $\rho $, i.e. $p_i \in  \mathrm{cl}[p]$
for all $p \in E-K$;

\item the only open and $\rho $-saturated neighborhood of every $p_i, \,
i=1,  \dots ,r$, is $E$;

\item the algebra $\mathcal{C}_{\rho }(E)$ of global, $\rho $-consistent
cross  sections of the sheaf $\mathcal{C}_E$ is isomorphic with $\mathbf{R}$.
\end{enumerate}
\end{Theorem}

\textbf{Proof} $(1) \Rightarrow (2).$ Let $U$ be an open and $\rho $%
-saturated neighborhood of $p_i$. Since $p_i \in \mathrm{cl}[p]$ for every $%
p \in E-K$, there exists a point $q \in [p]$ such that $q \in U$ ($p_i$ are
boundary points of the set $[p]$). But $U$ is $\rho $-saturated, therefore $%
p_i \in U\supset\ [q] = [p]$, and consequently $U=E$.

$(2)\Rightarrow (3)$. Let $E$ be the only open and $\rho $-saturated
neighborhood of $p_i$. Let us consider $\alpha \in \mathcal{C}_{\rho }(E)$,
and let us suppose that there exists a point $p\in E$, $p\neq p_i$, such
that $\alpha (p) \neq \alpha (p_i)$. Then there exists $\varepsilon > 0$, $%
\varepsilon = \frac{|\alpha(p)-\alpha(p_i)|}{3}$, such that the sets $%
(\alpha(p)-\varepsilon , \alpha(p) + \varepsilon )$ and $(\alpha(p_i)-%
\varepsilon , \alpha(p_i) + \varepsilon )$ are disjoint. In such a case, the
set $\alpha ^{-1}(\alpha (p_i - \varepsilon ), \alpha (p_i + \varepsilon ))$
is open, $\rho $-saturated, and disjoint with $\alpha ^{-1}(\alpha (p -
\varepsilon ), \alpha (p + \varepsilon ))$ which contradicts the assumption.

$(2) \Rightarrow (1)$. Let $p_i \in E$ be such that its unique open and $%
\rho $-saturated neighborhood is $E$, and let us suppose that $p_i \notin
\mathrm{cl}[p]$ for a certain $p \in E-K$. It follows that $E - \mathrm{cl}%
[p]$ is an open (as a completion of a closed set) and $\rho $-saturated
neighborhood of $p_i$ which contradicts the assumption.

$(3) \Rightarrow (2)$. Let $\alpha \in \mathcal{C}_{\rho }(E)$, and let $%
\alpha (p_i) = k =$ const. Since $\alpha $ is a constant function then for
every $\varepsilon > 0$, $\alpha ^{-1}(k-\varepsilon , k+\varepsilon) = E$
is an open and $\rho $-saturated neighborhood of $p_i$ and it does not
depend on $\varepsilon $. Sets of the form $\alpha ^{-1}(k-\varepsilon ,
k+\varepsilon) = E$ create a subbase of the topology of open and $\rho $%
-saturated sets of $E$. It follows that $E$ is the only open and $\rho $%
-saturated set containing $p_i$. $\diamond $

All three conditions of the above theorem are satisfied in the case of the
closed Friedman world model and Schwarzschild solution \cite{GRG94,Origin}.

In the following, for simplicity, we shall consider $E$ with one ``singular
point'' $\{p_0\}$, i.e., $K = \{p_0\}$. Analysis with a finite number of
``singular points'' goes in an analogous manner.

\section{SINGULARITIES IN NONCOMMUTATIVE REGIME}
\label{sec3}

\subsection{Preliminaries}

\label{ssec3.1} We put Schmidt's construction into the noncommutative
context in the following way. Let $\pi _{M}:E\rightarrow M$ be, as before,
the frame bundle over space-time $M$ with $G$ as its structural group
(typically, $G=SO_{0}(3,1)$), and let us remember that both $M$ and $E$ are locally
compact and smooth manifolds. Since $G$ is acting on $E$, $E\times
G\rightarrow E$, we can naturally define the groupoid $\Gamma _{1}=E\times G$%
, and the noncommutative algebra $\mathcal{A}_{1}=C_{c}^{\infty }(\Gamma
_{1},\mathbf{C})$ of smooth, compactly supported, complex valued functions
on $\Gamma _{1}$ with convolution as multiplication (both groupoid $\Gamma
_{1}$ and algebra $\mathcal{A}_{1}$ are described in \cite{JMP05,Pysiak06}).

Let us also consider the pair groupoid $\Gamma=\bigcup E_x \times E_x$.
Every $E_x,\, x \in M$, is an equivalence class of the equivalence relation $
\rho \subset E \times E$ which is $F$-regular. We assume that $F = G$. The
relation $\rho $ is defined in the following way: $p_1 \rho p_2$ iff there
exists $g \in G$ such that $p_2 = p_1g$. It can be demonstrated that the
groupoids $\Gamma_1$ and $\Gamma $ are isomorphic \cite{JMP05}, but the
groupoid $\Gamma $ suits better for analyzing singularities.

Let us define $\bar{E} = E \cup \{p_0\}$ and assume that the point $p_0$ remains
in close contact with all equivalence classes of the relation $\rho $. Let us also
extend the relation $\rho $ to $ \bar{E}$ as follows: $\bar{\rho } \subset \bar{E}
\times \bar{E}, \, (p_0, p_0) \in \bar{\rho }$. The relation $\bar{\rho }$ is $F$-regular with the
exception of the point $\{p_0\}$. Let us denote $\bar{E}/\bar{\rho }$ by $\bar{M}$.
We now have the groupoid $\bar{\Gamma } = \Gamma \cup \{(p_0,p_0)\}$ over $\bar{E}$.
Clearly, it is the groupoid of the equivalence relation $\bar{\rho }$.

We define the noncommutative algebra $\mathcal{A}$ on the groupoid $\Gamma $%
, $\mathcal{A}=(C_{c}^{\infty }(\Gamma ,\mathbf{C}),\ast )$, where
\textquotedblleft $\ast $\textquotedblright\ denotes convolution, and its
outer center $Z=\pi _{M}^{\ast }(\mathcal{C}(M))$ where $\mathcal{C}(M)$ is
the algebra of smooth functions on $M$. In the following we also
consider the subalgebra of $Z$ consisting of all bounded functions, $Z_b
\subset \pi _{M}^{*}(\mathcal{C}(M))$. Let us notice that the
algebra $Z_{b}$ contains all information necessary to reconstruct the
algebra $\mathcal{C}(M)$. Indeed, for every point $x \in M$ there exists a
neighborhood $U \ni x$ such that $\mathrm{cl}U$ is a compact set, and consequently
$f|\mathrm{cl}U$, where $f \in \mathcal{C}(M)$, is a bounded function. In such
a case, $g = f \cdot \phi $, where $\phi $ is a
bump function, is a smooth bounded function on $M$. One has
\[
f|V = g|V = \mathrm{id}_{\mathbf{R}} \circ g|V,
\]
which means that the differential structure $C^{\infty }(M)$ is generated by
$C^{\infty }_b(M) \simeq Z_b$. We also define $\bar{Z}=\pi _{\bar{M}}^*(%
\mathcal{C}(\bar{M}))$ containing only constant functions on
$\bar{E}$ (which are automatically bounded). We will understand
that $\bar{Z} \subset Z_b$.

The outer center $Z$ acts on the algebra $\mathcal{A}$, $\alpha :Z\times
\mathcal{A}\rightarrow \mathcal{A}$, in the following way
\[
\alpha (f,a)(p_1, p_2) = f(p_1)a(p_1,p_2),
\]%
$a \in \mathcal{A}, \, f \in Z$.

The space-time geometry is encoded in the outer center $Z$. Since singularities are properties of space-time (in the sense that it is a given space-time that is singular), it seems reasonable to consider the extended algebra $\mathcal{A}_1 = \mathcal{A} \oplus Z$. It is a noncommutative algebra with the multiplication, also denoted by $*$, given by the following formula
\[
(a+f)*(b+g) = a*b + \alpha (f,b) + \alpha (g,a) + f \cdot g,
\]
where $a,b \in \mathcal{A},f, g \in Z$. It is a central extension of the algebra $\mathcal{A}$ by $Z$. Now, $Z$ is a center (in the usual sense) of the algebra $\mathcal{A}_1$. We shall also consider the algebra $\mathcal{A}_{1b} = \mathcal{A} \oplus Z_b$.

Let us now construct the algebra $\bar{\mathcal{A}}_1$ on the groupoid $\bar{\Gamma }$. We (smoothly) extend functions from $\mathcal{A}$ through zero, i.e., for $a \in \mathcal{A}$ we have $\bar{a}(p_0, p_0) = 0$, and $\bar{a}|\Gamma = a$. This guarantees that the convolution operation is well defined. It is obvious that the algebra $Z$ reduces to $\bar{Z}$. The latter algebra consists only of constant functions (if the smoothness of the extended functions is assumed). Consequently, the algebra on $\bar{\Gamma }$ is of the form $\bar{\mathcal{A}}_1 = \mathcal{A} \oplus \bar{Z}$.
We shall also write $\bar{\mathcal{A}}_1 \subset \mathcal{A}_1$, and $\bar{\mathcal{A}}_1 \subset \mathcal{A}_{1b}$. The algebra $\bar{\mathcal{A}}_1$ is a noncommutative algebra corresponding to space-time with a malicious singularity.

Now, we construct the representation of the algebra $\mathcal{A}_{1b}$ in the
Hilbert space $\mathcal{H}_{p}=L^{2}(\Gamma ^{p})$,
\[
(\pi _{p}(a)\psi )(p_{1},p)=\int_{E_{x}}a(p_{1},p_{2})\psi (p_{2},p)dp_{2}
\]%
for $a\in \mathcal{A},\,\psi \in \mathcal{H}_{p},\,p,p_{1},p_{2}\in E_x$ with $%
x=\pi _{M}(p)$. For $f\in Z_{b}$ we prolong this representation in the
following way
\[
(\pi _{p}(f)\psi )(p_{1},p)=f(p_{1})\psi (p_{1},p),
\]
and
\[
\pi_p(a+f) = \pi_p(a) + \pi_p(f)
\]
for $a+b \in \mathcal{A}_{1b}$. It can be easily seen that $\pi_p$ is indeed a representation of the algebra $\mathcal{A}_{1b}$ Since $\bar{\mathcal{A}}_1 \subset \mathcal{A}_{1b}$, $\pi_p $ is also a representation of the algebra $\bar{\mathcal{A}}_1$. For $\bar{f} = \lambda = \mathrm{const} \in \bar{Z}$ we have
\[
(\pi _{p}(\bar{f})\psi )(p_{1},p)=\lambda \psi (p_{1},p).
\]

\subsection{Probabilistic Properties}

\label{ssec3.2}

It is vital to notice that every $a\in \mathcal{A}$ generates a random
operator $r_a=(\pi_p(a))_{p\in E}$ acting on a collection of Hilbert spaces $%
\{\mathcal{H}_p\}_{p \in E}$ where $\mathcal{H}_ p=L^2(\Gamma_p)$; here $%
\Gamma^p$ denotes all elements of $\Gamma $ with $p$ as a second element of
the pair. Let us first remember that an operator valued function $E \ni p \mapsto r(p) \in \mathcal{B}(\mathcal{H})_p$ is a \emph{random
operator\/} if it satisfies the following conditions \cite[p. 50-53]{Connes}:

(1) If $\xi_p,\eta_p \in \mathcal{H}_p$ then the function $ E\rightarrow
\mathbf{C}$, given by $E\ni p\mapsto (r(p) \xi_p,\eta_p)$, $a\in \mathcal{A}$%
, is measurable with respect to the usual manifold measure on $E$.

(2) The operator $r$ is \emph{bounded}, i.e., $||r|| < \infty$, $%
||r||=\,\mathrm{e}\mathrm{s} \mathrm{s}\,\mathrm{s}\mathrm{u} \mathrm{p}
||r(p)||$, where ``ess sup'' denotes supremum modulo zero measure sets
(essential supremum).

Let us denote by $\mathcal{R}$ the space of equivalence classes (modulo equality almost everywhere) of operator valued functions satisfying the above two conditions. $\mathcal{R}$ is an algebra with the multiplication  $[r_1(p)][r_2(p)] = [r_1(p)r_2(p)]$ and the involution  $[r(p)]^* = [r(p)^*]$.

Operators $r_a=(\pi_p(a)),p\in E$, satisfy conditions  (1) and (2) (for details see
\cite{JMP05,Pysiak06}).

We define the algebra $\mathcal{M}_{0}$ of equivalence classes (modulo
equality almost everywhere) of random operators $r_{a},a\in \mathcal{A}$, and the commutant of $\mathcal{M}_0$ as
\[
\mathcal{M}_0' = \{[r] \in \mathcal{R}: [r][r_a] = [r_a][r] \mathrm{for\; every} \; [r_a] \in \mathcal{M}_0\}.
\]
Now, we complete $\mathcal{M}_0$ to the von Neumann algebra
$\mathcal{M}= \mathcal{M}_0''$ where
$\mathcal{M}_{0}^{\prime \prime }$ denotes the double commutant of
$ \mathcal{M}_{0}$, i.e., the commutant of $\mathcal{M}_0'$ in
$\mathcal{R}$. The algebra $\mathcal{M}$ is called \emph{von
Neumann algebra of the groupoid} $\Gamma $. It can be shown
that $\mathcal{M}$ is isomorphic to the algebra $L^{\infty }(M, \mathcal{B}%
(L^2(G)))$ \cite{Pysiak06}.

Let us denote by $\mathcal{M}_{01}$ the algebra of equivalence classes of $r_{a_1} = (\pi_p(a_1))_{p\in E},\, a_1 \in \mathcal{A}_1$. Let us also notice that $\mathcal{M} = L^{\infty }(M, \mathcal{B} (L^2(G)))$ implies that
$\mathcal{M}_{01}'' = \mathcal{M}$.

Now, we consider $\bar{E} = E\cup \{p_0\}$. Since $\bar{E}$ is a manifold,
it carries a Lebesgue measure (the set $\{p_0\}$ is of zero measure).
The family of Hilbert spaces becomes
\[
\{\mathcal{H}_{p}\}_{p \in \bar{E}} = \{\mathcal{H}_p\}_{p\in E} \cup \{%
\mathcal{H}_{p_0}\}.
\]
The Hilbert space $\mathcal{H}_{p} = L^2(\Gamma^{p})$ is like before, and $%
\mathcal{H}_{p_0} = L^2(\Gamma^{p_0}) \simeq \mathbf{C}$. Since $\Gamma^{p_0}
$ denotes elements of $\Gamma $ that end at $p_0$, in this case we have $\Gamma^{p_0}=\{(p_0,p_0)\}$, and square integrable functions at a point give
constants.

Let $\bar{a} \in \bar{\mathcal{A}}_1, \, \psi \in \mathcal{H}_p,\, p \in E$, where $\bar{a} = a + \lambda, \, a \in \mathcal{A}, \, \lambda \in \bar{Z}$. In this case,
\[
\pi_p(\bar{a}) = \pi_p(a) + \pi_p(\lambda ).
\]
Therefore, everything happens as in the regular part, but
\[
(\pi_{p_0}(\bar{a})\psi)(p_0,p_0) = \lambda \psi(p_0,p_0)
\]
where $\psi(p_0,p_0)\in \mathbf{C}$.

Now, we define the algebra $\bar{\mathcal{M}}_0$ of random operators of the
type $r_{\bar{a}}, \, a \in \bar{A}_1$.
It is obvious that $\mathcal{M}_0 \subset \bar{\mathcal{M}}_0 \subset \mathcal{M}_{01}$, and we obtain the interesting result
\[
(\mathcal{M}_0)^{\prime\prime}= (\bar{\mathcal{M}_0})^{\prime\prime}=
\mathcal{M}.
\]

Let us notice that $\mathcal{M} \simeq \bar{\mathcal{M}}$. $\mathcal{M}
_{0}$ is the algebra of random operators before the \textquotedblleft
singular point\textquotedblright\ has been attached, and $\bar{\mathcal{M}}%
_{0}$ is the algebra of random operators after the \textquotedblleft
singular point\textquotedblright\ has been attached. If we try to complete
these two algebras to the von Neumann algebra, we obtain the same von
Neumann algebra $\mathcal{M}$. This means that from the probabilistic point
of view the \textquotedblleft singular point\textquotedblright\ is
irrelevant. Probabilistic aspects of the algebra $\mathcal{M}$ are studied
in \cite{Random}.

\section{INTERPRETATION}

\label{sec4} Mathematical tools based on the concept of differential
manifold are typical for analyzing singularities in the macroscopic context
of general relativity. To these tools belong: g-incompleteness of
space-time, its conformal and causal structure, various space-time
boundaries... From the macroscopic point of view, the existence of
singularities, especially of strong ones, is a signal that something is
going on wrong. To elucidate the nature of this catastrophe we have used the
tools of differential spaces and the tools of structured spaces. The
advantage of these tools with respect to more traditional methods is that,
with the help of them, singularities can be regarded as ``internal
elements'' of a differential (resp. structured) space, and not necessarily
as its ideal points (forming some kind of boundary). Owing to the fact that
the differential structure of structured spaces consists of a sheaf of
algebras, no pathologies occur when the sheaf is restricted to the regular
(nonsingular) part of a given structured space. Only when the sheaf is
extended to singularities, it collapses to constant functions. Theorem 1 of
Sec. \ref{sec2} elucidates the nature of catastrophe which then occurs.

Such a catastrophe can be studied with the help of noncommutative geometry.
Geometric properties of a noncommutative space are encoded in a certain
noncommutative algebra. In our case, it is a von Neumann algebra of random
operators and, as we have seen, this means that, from the probabilistic
point of view, singularities, in this setting, are irrelevant.

This result seems very interesting from the physical point of view. Often
the question is asked of whether the singularity will persist when finally
the correct quantum gravity theory is discovered (the question is mainly
addressed to strong singularities of the Big Bang type). Usually, either
``yes'', or ``no'' answers are given to this question. If we agree that
noncommutative geometry will somehow be engaged into the future quantum
gravity theory then the problem is not of whether do singularities exist on
the fundamental level or not, but rather of whether they are relevant or
not. And our answer is that they are irrelevant. Only when one goes to the
macroscopic level, one must take a ratio of a given algebra by an
equivalence relation (one could say that some sort of ``averaging'' is
occurring) and then the singularities appear.

\end{document}